\documentclass[journal=jacsat,manuscript=article]{achemso}
\usepackage{amsmath}
\usepackage{amstext}
\usepackage{amsfonts}
\usepackage{graphicx}
\usepackage{xcolor}

\usepackage{natbib}
\usepackage[pdftex]{hyperref}
\hypersetup{
    unicode=false,          
    pdftoolbar=true,        
    pdfmenubar=true,        
    pdffitwindow=false,     
    pdfstartview={FitH},    
    colorlinks=false,        
}

\title{Exact analytical algorithm for
solvent accessible surface area and derivatives in implicit solvent molecular simulations on GPUs}

\author{Xin Cao}
\affiliation{Department of Applied Mathematics and Statistics, Stony Brook University, Stony Brook, NY 11794, United States}
\alsoaffiliation{Laufer Center for Physical and Quantitative Biology, Stony Brook University, Stony Brook, NY 11794, United States}

\author{Michelle~H. Hummel}
\affiliation{Sandia National Laboratories, Albuquerque, NM 87123, United States }

\author{Yuzhang Wang}
\affiliation{Department of Chemistry, Stony Brook University, Stony Brook, NY 11794, United States}

\author{Carlos Simmerling}
\affiliation{Department of Chemistry, Stony Brook University, Stony Brook, NY 11794, United States}
\alsoaffiliation{Laufer Center for Physical and Quantitative Biology, Stony Brook University, Stony Brook, NY 11794, United States}

\author{Evangelos~A. Coutsias}
\email{evangelos.coutsias@stonybrook.edu}
\affiliation{Department of Applied Mathematics and Statistics, Stony Brook University, Stony Brook, NY 11794, United States}
\alsoaffiliation{Laufer Center for Physical and Quantitative Biology, Stony Brook University, Stony Brook, NY 11794, United States}

\begin{document}

\begin{abstract}
In this paper, we present dSASA (differentiable SASA), an exact geometric method to calculate solvent accessible surface area (SASA) analytically along with atomic derivatives on GPUs. The atoms in a molecule are first assigned to tetrahedra in groups of four atoms by Delaunay tetrahedrization adapted for efficient GPU implementation and the SASA values for atoms and molecules are calculated based on the tetrahedrization information and inclusion-exclusion method. The SASA values from the numerical icosahedral-based method can be reproduced with more than 98\% accuracy for both proteins and RNAs. Having been implemented on GPUs and incorporated into the software Amber, we can apply dSASA to implicit solvent molecular dynamics simulations with inclusion of this nonpolar term. The current GPU version of GB/SA simulations has been accelerated up to nearly 20-fold compared to the CPU version, outperforming LCPO, a commonly used, fast algorithm for calculating SASA, as the system size increases. While we focus on the accuracy of the SASA calculations for proteins and nucleic acids, we also demonstrate stable GB/SA MD mini-protein simulations.

\end{abstract}


\maketitle

\section{Introduction}
An accurate description of the solvent environment is essential for biomolecular modeling, as biological machines are functioning in an aqueous environment. The solute-solvent interactions and the rearrangement of water molecules induce the change of molecular shapes and the solvation free energy $\Delta G_{sol}$ \cite{fennell11}. Including water molecules in explicit solvent simulations produces more accurate simulation results at the cost of redundant calculation for the pairwise interactions between water molecules, while the friction induced by collisions with water molecules can further slow down the conformational sampling of the solute. 
In implicit solvent models, the solvent is treated as a continuum and the system is simulated without explicit water molecules. The major advantage of implicit solvent models is the ability to explore the conformational space more efficiently, which can find applications in protein folding studies and structure prediction \cite{nguyen14, perez16}, solvation free energy calculation using Poisson-Boltzmann surface area (PBSA) \cite{stikoff94} or generalized Born surface area (GBSA) \cite{still90}, and binding free energy calculation using molecular mechanics Poisson-Boltzmann surface area (MM-PBSA) \cite{kollman00}. 

In implicit solvent modeling, the solvation free energy can be decomposed into polar and nonpolar contributions 
\cite{levy03} and the expression is
$$ 
\Delta G_{sol} = \Delta G_{pol} + \Delta G_{cav} + \Delta G_{vdW} = \Delta G_{pol} + \Delta G_{np},
$$ 
where the polar term $\Delta G_{pol}$ gives the difference between the work of uncharging the solute in vacuum and the work of charging the solute in solvent, $\Delta G_{cav}$ denotes the accommodation of cavity in the solvent for solute and $\Delta G_{vdW}$ represents the van der Waals interactions between solute and solvent.
The Poisson-Boltzmann (PB) method describes the electrostatic potential in solution for a given set of boundary conditions and determines the effect of electrostatic interactions on the molecules \cite{honig95} and can be used to compute $\Delta G_{pol}$. The $\Delta G_{cav}$ and $\Delta G_{vdW}$ are usually combined into one nonpolar term $\Delta G_{np}$. It is considered to be proportional to the number of atoms in the solute having direct contact with solvent molecules and can be estimated in terms of the solvent accessible surface area (SASA) \cite{eisenberg86}. Although it has been pointed out that incorporation of the volume term into the computation can provide a more complete description of the nonpolar contribution \cite{wagoner06, allison11}, the surface area based methods can still have good performance in the prediction of the native-like conformations of proteins and the estimation of ligand-binding affinities \cite{genheden15}.

In implicit solvent molecular dynamics (MD) simulations, most of the effort has been devoted to the development of the dominant polar part. Due to the complexity of solving the PB equation, the generalized Born (GB) method \cite{still90} was proposed to approximate the solution of the PB equation with simple functional forms. The recent improvement of fast GB models for proteins \cite{onufriev04, mongan07, nguyen13} and nucleic acids \cite{nguyen15} and its implementation on graphics processing units (GPUs) \cite{case17} helped GB achieve popularity in recent years. 
The main goal of a GB model is to achieve agreement of electrostatic free energies with the PB approach, which can be achieved by parametrizing the GB screening parameters to provide a better match of the effective radii to those from the reference method.
In particular, the pairwise decomposition version of GB \cite{hawkins95} is ideal for implementation on GPU with parallel computing. 
Although the performance of GB is improving, poor folding stabilities were observed in protein folding studies \cite{nguyen14}, structure predictions \cite{perez16} and de novo designed peptides with high helicity \cite{lang22}. As hypothesized in \cite{nguyen14}, the neglected nonpolar solvation term might result in instability; this was supported by improved fold stability when an approximate SASA-based nonpolar term was added to the same GB model\cite{Huang18}. As shown in that paper, the inclusion of the nonpolar term in the GB/SA MD simulations can produce more stable trajectories and better simulated melting temperatures than comparable simulations with identical setup other than the SASA term.

There exist several methods to calculate the molecular surface area on CPUs and GPUs. However, only a few of them can be applied to MD simulations because most of them lack reliable derivatives for atoms. 
The solvent accessible surface was proposed by Lee and Richards \cite{lee71}. A numerical implementation by Connolly \cite{connolly83} is computationally expensive and lacks analytical derivatives. 
The derivatives are unreliable or unavailable in other methods, such as grid based, neighbor-counting or machine learning \cite{durham09}. The approximation methods on GPU have similar issues \cite{juba08,dynerman09}. 
In the CHARMM software, GBMV2/SA \cite{lee17} can reproduce the molecular surface and has been implemented on GPUs \cite{gong20}.
However, the computation is based on grid points so lower resolution of the grids may generate less reliable results while higher resolution results in slower simulation speed. 
The first available method for approximating surface area along with derivatives in Amber is LCPO \cite{weiser99} which is able to estimate SASA based on the neighbors of atoms. As the MD simulations in Amber are now primarily on GPUs \cite{gotz12, case17}, the data transfer between CPU and GPU deteriorates the performance because LCPO is implemented only on CPUs. Recently, another approximate pairwise method with a faster speed (up to ~30 times faster), called pwSASA, was proposed and it achieved a comparable accuracy to LCPO according to the testing of proteins \cite{Huang18}. The main purpose of pwSASA was to investigate the impact of the nonpolar term on the MD simulations of proteins with a simple and fast pairwise approximation suitable for running entirely on GPUs. The simple form of the functions provided faster simulation speed and did prove the significance of nonpolar term in the simulations. However, the computed SASA values had various correlation coefficients from 0.6 to 0.9 compared with the numerical computation for a number of proteins, which is partly because the higher order interactions were ignored. Furthermore, pwSASA is highly empirical, and the parameters were trained only for the specific local atomic environments founds in amino acids (such as hybridization and the identity of bound atoms). Thus in addition to relatively low accuracy on proteins, pwSASA cannot be used in simulations of more diverse systems such as those including modified amino acids, small molecule ligands, nucleic acids, or glycans, further limiting its suitability.
 \color{black}
 
As every atom in a molecule can be represented by a sphere,
the surface areas have also been studied by computational geometry methods. One reliable and accurate method was developed based on the Alpha Complex construction \cite{edelsbrunner95} by which the geometric descriptors of molecules can be computed through union of balls. The Alpha Complex is obtained with the atomic coordinates and radii information and can be applied to decompose
the space into small segments through which the surface areas are computed by using Gauss-Bonnet theorem \cite{cazals11} or inclusion-exclusion formulas \cite{bryant04, hummel14, hummel18}. Recently, the extension 
to other geometric measures, such as volumes, mean and Gaussian curvatures, was proposed \cite{koehl23} in which the parallel implementation can achieve speed improvement over the serial version. 
However, the full implementation of such method on GPUs is still unavailable because each step in the method requires substantial effort to utilize the parallel computing property to improve performance.

Here we present dSASA, a new method to calculate SASA and its derivatives with respect to atomic coordinates using Alpha Complex theory and inclusion-exclusion method. 
This fully analytical and accurate algorithm has been implemented on GPUs and incorporated in Amber software. The basic background and the procedures employed in the method will be first shown. 
The results for the assessment of the method are next given:
the accuracy of estimation is examined by several protein systems with diverse topologies, 
the speed of the GPU version in GB/SA simulations is compared with the CPU version and other methods, and
the performance is demonstrated by including this nonpolar term in GB/SA simulations on proteins. However, we caution readers that good practice in force field development involves isolation and validation of the individual components as done here. Deviations from experimental behavior can arise from inaccuracy in any of the multiple components (solute force field, polar solvation term, and nonpolar term). Likewise, a good match to experiment can be obtained from fortuitous cancellation of error. For this reason, we show that the dSASA term performs well in MD simulations, which verifies the efficient algorithm implementation, but our main focus remains on validating the SASA accuracy.

\section{Methods}\label{sasa:methods}
\subsection{Theory and estimations for nonpolar solvation}
In the implicit solvent model, the nonpolar solvation term is usually SASA-based \cite{still90}. 
The free energy is proposed to be proportional to SASA with a surface tension parameter $(\gamma)$:
$\Delta G_{np} = \gamma SASA.$ The surface tension is assumed to be identical for all atoms.

The relatively accurate SASA values can be calculated by ICOSA numerical method \cite{shrake73} ($gbsa = 2$ in Amber), in which starting from an icosahedron a water probe with radius 1.4~\AA~is recursively rolled on the van der Waals surface of the molecule. 
The implementation of this method in current Amber MD simulations is unavailable because of the absence of atomic forces. The Linear Combinations of Pairwise Overlaps (LCPO; $gbsa=1$ in Amber) is the first stable algorithm used in GB/SA MD simulations in Amber. 
The neighbor list of a central atom is used to subtract the pairwise overlapping from its isolated sphere areas, which results in the computation complexity $O(N^2)$, where $N$ is the total number of solute atoms, so it will slow down quadratically as the size of the molecule increases. In the GPU version of GB/SA MD with this method, the data transfer between CPU and GPU decelerates the simulations as its implementation is CPU/GPU hybrid. 
The pairwise approximation of SASA (pwSASA; $gbsa=3$ in Amber) \cite{Huang18} is the recently developed algorithm to calculate SASA that is suitable for implementing on GPUs. 
The idea is similar to LCPO but only considers the interactions between two atoms, which achieves a faster speed but results in an approximate SASA. The parameters for the atom types are trained exclusively on standaard amino acids, so the application to other systems, such as nucleic acids or small molecules, is unavailable.

\subsection{dSASA: an analytical method for SASA} 

\begin{figure}
\centering
\includegraphics[scale=0.35]{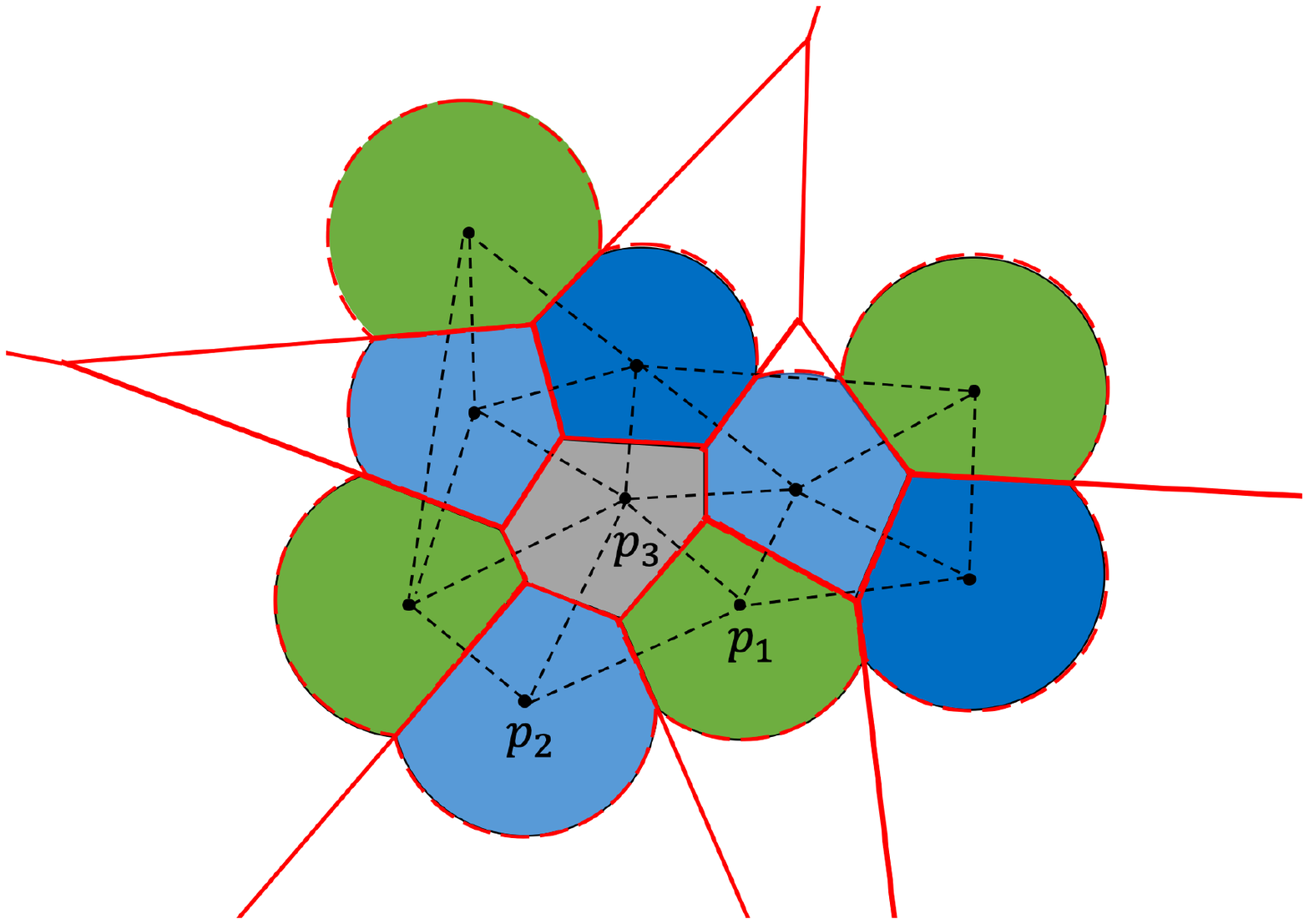}
\caption[Depiction of the SASA method in 2D.]{Depiction of the method in 2D. Every solid circle is the expanded atom including the water probe, black points show the center of atoms. The dashed black lines denote part of the tetrahedrization of the molecule. The atoms in green and blue (such as $p_1, p_2$) are exterior, grey atom $p_3$ is interior. The boundary of Laguerre cells of atoms are shown in solid red lines, and exterior boundary of Laguerre cells of molecule in solvent is represented by the dashed red curves which correspond to SASA.}
\label{method:lag1}
\end{figure}

Here we introduce {\em dSASA}, a geometric method ($gbsa=4$ in Amber) to compute SASA. As has been explored in \cite{weiser99, vasil02}, to be consistent with other methods, we assume that the heavy atoms in molecules can have a good approximation to the molecular SASA, reducing computational cost substantially. 
The atomic radii for four common elements (C, 1.7 \AA; O, 1.5 \AA; N, 1.55 \AA; S, 1.8 \AA) are used in ICOSA, pwSASA and dSASA, while LCPO is using different radii (C, 1.7 \AA; O, 1.6 \AA; N, 1.65 \AA; S, 1.9 \AA). Every atom can be represented by a weighted point $p_i^{\prime} = (p_i, d_i)$ in the point set $\mathbb{A} \subset \mathbb{R}^3 \times \mathbb{R}$, which contains the 3D coordinates $p_i$ in space and one point weight $d_i$. The radius of water probe is set to 1.4~\AA, and the weight $d_i = (\text{atom\_radius}_i+1.4)^2$. Here we present the key steps, while a complete description of the analytical development of the method is given in \cite{hummel14,hummel18}.
The method has three main steps and the depiction of the method is shown in Figure~\ref{method:lag1}. \\
{\bf Step 1.} Calculate the 3D weighted Delaunay Tetrahedrization (wDT) of all points through which the points are assigned into tetrahedra. This step has computation complexity $O(N \log N)$ and is the dominating part of the method. \\
{\bf Step 2.} Create and classify the dual complex $\mathcal{C}$ which denotes the possible interactions among close atoms. The complex $\mathcal{C}$ contains the lists of simplices, such as the vertices (the atoms), edges (overlap between two atoms), triangles (overlap among three atoms) and tetrahedra (overlap among four atoms) from the tetrahedrization. These simplices are then filtered and classified into interior and exterior (denoted by $\partial \mathcal{C}$) based on their connection information. \\
{\bf Step 3.} Compute the surface areas using Laguerre intersection cells and inclusion-exclusion method based on the exterior simplices $\partial \mathcal{C}$.  

In the following, we show the process to compute the atomic SASA based on the exterior simplices $\partial \mathcal{C}$ because only the atoms having direct contact with solvent contribute to the surface areas. The simplices have been classified into interior and exterior at Step 2.
The Laguerre diagram of the point set is conjugate to the wDT. The space can be decomposed into cells by the conjugated Laguerre diagram as shown in Figure~\ref{method:lag1}. 
The Laguerre cell for a weighted point (atom) $p_i^{\prime}$ is defined by 
$$
L_i= \{ x \in \mathbb{R}^3 : | p_i - x |^2 - d_i \le | p_j - x |^2 - d_j, \; \forall p_j^{\prime} \in \mathbb{A} \}.
$$
Every weighted point can also be treated as a ball $B_i = \{x \in \mathbb{R}^3: |p_i - x|^2 - d_i \le 0 \}$, and the union of these balls $B = \cup B_i$ forms the space filling model of $\mathbb{A}$.
As shown in Figure~\ref{method:lag1}, the Laguerre cells of exterior atoms can be unbounded or larger than the real size of the atoms, 
then a more realistic representation for an atom is given by the Laguerre-intersection cell, the intersection part of the Laguerre cell and the ball: $LI_i = L_i \cap \ B_i.$
\begin{figure}
\centering
\includegraphics[scale=0.4]{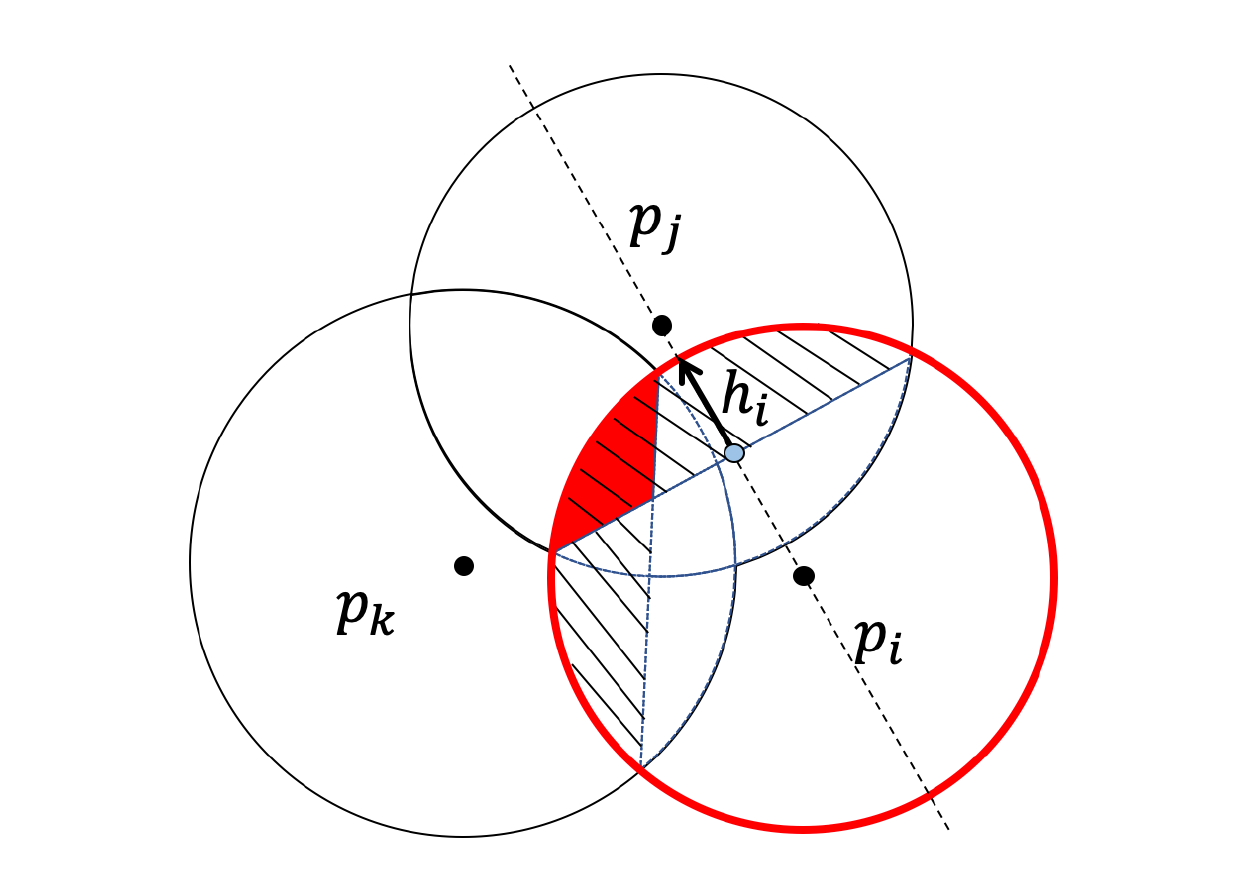}
\caption{Inclusion-exclusion process of $p_i$. The red circle denotes the standalone ball $S_i$ of $p_i$; the shaded region indicates the intersection $S_{ij}$ and $S_{ik}$ with $p_j, p_k$; the small red region is the intersection of three atoms $S_{ijk}$.}
\label{method:inexclu3}
\end{figure}
The boundary area of the Laguerre intersection part $LI_i$ of an exterior atom is the SASA value $\mathcal{S}^i = surf(LI_i)$. 

Suppose there are $k$ points in a subset $T \subset \mathbb{A}$. The number of points in $T$ is $|T| = k$ and the centers of points are denoted by $T^{\prime}$. The convex hull ($conv$) of points in $T^{\prime}$ is written as $\sigma_{T}$, and the term $S_{T}$ is the surface area of the intersection of the balls in $T$. By applying the inclusion-exclusion method shown in Figure~\ref{method:inexclu3}, the SASA of the molecule is
$$ 
\mathcal{S} = \sum_{\sigma_{T} \in \partial \mathcal{C}} (-1)^{k+1} c_{T} S_{T}, \; k= 1,2,3.
$$
The coefficients $c_{T}$ are given by the corresponding simplices. When $|T|=1$, i.e. a vertex $v_i$: $c_{T} = \Omega_{T}$ is the fraction of the ball $i$ outside the tetrahedra in $\mathcal{C}$. $\Omega_{T}$ is the normalized outer solid angle subtended by the union of tetrahedra in $\mathcal{C}$ containing $v_i$. When $|T|=2$, i.e. an edge $e_{ij}$: $c_{T} = \Phi_{T}$ is normalized outer dihedral angle of the union of tetrahedra in $\mathcal{C}$ which contain the edge $e_{ij}$. When $|T|=3$, i.e. a triangle $t_{ijk}$: $c_{T} = 1$ or $0.5$ is the fraction of $S_{T}$ that is outside the union of tetrahedra in $\mathcal{C}$. Here $v_i = p_i, e_{ij} = conv(\{p_i, p_j \})$, and $t_{ijk} = conv(\{p_i, p_j, p_k \})$. 

The contribution of individual atom $p_i^{\prime}$ to the molecular surface area is
\begin{align}\label{eqs:sainexclu}
\mathcal{S}^i =  \Omega_{i} S_{i}^{(i)} - \sum_{e_{ij} \in \partial \mathcal{C} } \Phi_{ij} S_{ij}^{(i)} + \sum_{t_{ijk} \in  \partial \mathcal{C} } c_{ijk} S_{ijk}^{(i)},
\end{align}
where $S_{T}^{(i)}$ is the contribution of $S_{T}$ to $\mathcal{S}^i$, and $\sum_{i=1}^{n} \mathcal{S}^i = \mathcal{S}$.

Then the corresponding derivative for the points
can be obtained on the basis of the atomic surface area with respect to the atomic coordinates in the molecule as  
\begin{align}\label{eqs:inexclu}
\nabla \mathcal{S}^i = \nabla  \Omega_{i} S_{i}^{(i)} - \sum_{e_{ij} \in \partial \mathcal{C} } (\nabla  \Phi_{ij} S_{ij}^{(i)} + \Phi_{ij} \nabla S_{ij}^{(i)}) + \sum_{t_{ijk} \in  \partial \mathcal{C} } c_{ijk} \nabla S_{ijk}^{(i)}.
\end{align}

\begin{figure}[ht]
\centering
\includegraphics[width=4.5in]{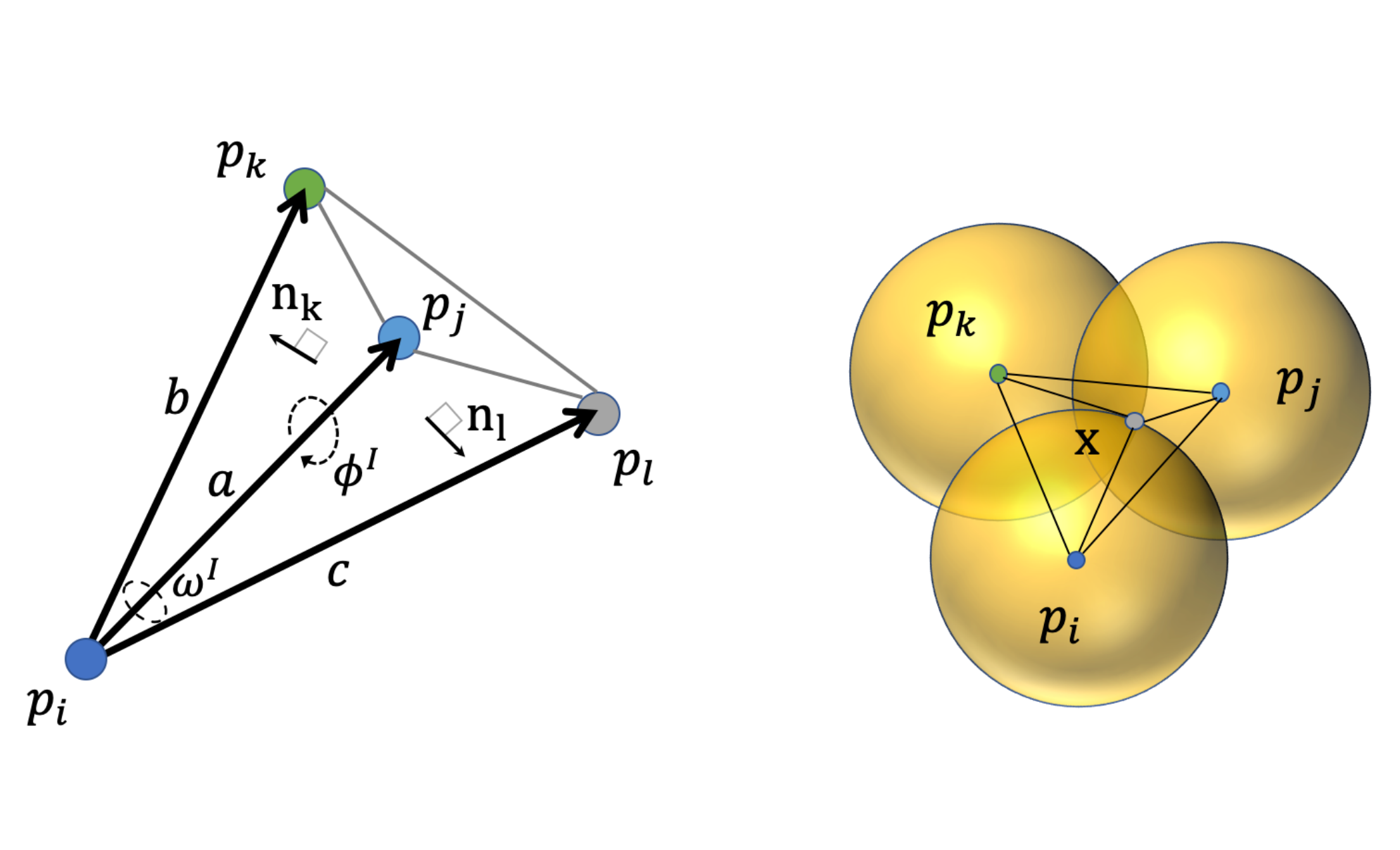}
\caption[Resulting tetrahedron from intersecting atoms.]{Left: Tetrahedron formed by points $p_i, p_j, p_k, p_l$. Right: Center $\bf{x}$ of the intersection of three balls and the resulting tetrahedron.}
\label{method:inters23}
\end{figure}

The terms in the equations are calculated as follows. When $|T|=1$, $S_{T}^{(i)} = 4 \pi d_i$. Let $\mathcal{I}_1$ be the set of tetrahedra in $\mathcal{C}$ to which the point $p_i^{\prime}$ belongs to. For $\zeta_{I} \in \mathcal{I}_1$ define $\omega^{I}$ as the normalized inner solid angle subtended by the tetrahedron $\zeta_{I}$ from the point $p_i^{\prime}$. Then 
\begin{align} \label{eqs:vertex}
\Omega_{T} = 1- \sum_{\zeta_{I} \in \mathcal{I}_1} \omega_{T}^{I}, \; \text{with the derivative } 
\nabla \Omega_{T} = - \sum_{\zeta_{I} \in \mathcal{I}_1} \nabla \omega_{T}^{I}.
\end{align}
The normalized inner solid angle, $\omega^{I}$, of a tetrahedron $p_i p_j p_k p_l$ subtended by the vectors $\bf{a}$ = $p_j- p_i$, $\bf{b}$ = $p_k- p_i$, and $\bf{c}$ = $p_l- p_i$ is shown in the left of Figure~\ref{method:inters23} and given by the equation
$$
\omega^{I} = \frac{1}{2\pi} \arctan \left(  \frac{|\mathbf{a} \cdot (\mathbf{b} \times \mathbf{c})|}{abc + (\mathbf{a} \cdot \mathbf{b})c + (\mathbf{a} \cdot \mathbf{c})b + (\mathbf{b} \cdot \mathbf{c})a }  \right),
$$
where $a = |\bf{a}|$ is the length of $\bf{a}$ and likewise for $b$ and $c$.

When $|T|=2$, the intersection of two atoms $p_i^{\prime}, p_j^{\prime}$ is shown in Figure~\ref{method:inexclu3}. $S_{T}^{(i)} = 2\pi \sqrt{d_i} h_i$. Let $\mathcal{I}_2$ be the set of tetrahedra in $\mathcal{C}$ containing the edge $\sigma_{T}$. For $\zeta_{I} \in \mathcal{I}_2$ define $\phi^{I}$ as the normalized inner dihedral angle of $\zeta_{I}$ along $\sigma_{T}$. Then  
$$
 \Phi_{T}  = 1- \sum_{\zeta_{I} \in \mathcal{I}_2} \phi_{T}^I.
$$
The normalized inner dihedral angle between planes $p_{i} p_{j} p_{k}$ and $p_{i} p_{j} p_{l}$ with normals $\bf{n}_k$ and $\bf{n}_l$ is shown in the left of Figure~\ref{method:inters23} and given by
\begin{align} \label{eqs:phiedge}
\phi^{I} = \frac{\arccos(\bf{n}_k \cdot \bf{n}_l)}{2\pi}, \text{ with } \nabla \phi^{I} = \frac{-\nabla (\bf{n}_k \cdot \bf{n}_l)}{2\pi \sqrt{1- (\bf{n}_k \cdot \bf{n}_l)^2}}.
\end{align}
Then the derivative terms for Eq.\eqref{eqs:inexclu} are
\begin{align} \label{eqs:edge}
\nabla S_{T}^{(i)} = 2\pi \sqrt{d_i} \nabla h_i, \;  \nabla  \Phi_{T}  = - \sum_{\zeta_{I} \in \mathcal{I}_2} \nabla \phi_{T}^I.
\end{align}

When $|T|=3$, consider $T=\{ p_i^{\prime}, p_j^{\prime}, p_k^{\prime}\}$. The surface area of the common intersection of three balls can be written as a weighted sum of the surface area of the single and double intersections. If $p_i^{\prime}, p_j^{\prime}, p_k^{\prime}$ have a non-empty intersection then there are two points in common with the surfaces of all three balls, shown in the right of Figure~\ref{method:inters23}. Denote one of the two points $\bf{x}$, define $p_x^{\prime} = (\bf{x},$ $0) \in \mathbb{R}^{3} \times \mathbb{R}$, and let $T_x = \{ p_i^{\prime}, p_j^{\prime}, p_k^{\prime}, p_x^{\prime} \}$. Let $S_2^x$ be the set of edges defined by $\sigma_{T_x}$ and $S_1^x$ the set of vertices in $\sigma_{T_x}$. The atomic surface area of $p_i^{\prime}$ from the intersection of $p_i^{\prime}, p_j^{\prime}, p_k^{\prime}$ is given by
\begin{align} \label{eqs:trianglev}
\frac{1}{2} \mathcal{S}^{(i)}_{T} =  \Phi_{ij}^{x} S^{(i)_x}_{ij} + \Phi_{ik}^{x} S^{(i)_{x}}_{ik}  - \Omega_i^{x} S_{i}^{x},
\end{align}
where $\Phi_{ij}^{x}$ is the normalized dihedral angle of $\sigma_{T_x}$ along the edge $\sigma_{ij}$, $\Omega_i^{x}$ is the normalized solid angle of $\sigma_{T_x}$ subtended from $p_i$ and similarly for other combinations $i,j,k$.
The derivative of this term is
\begin{align} \label{eqs:triangle}
\frac{1}{2} \nabla \mathcal{S}^{(i)}_{T} = \nabla \Phi_{ij}^{x} S^{(i)_{x}}_{ij} + \Phi_{ij}^{x}  \nabla S^{(i)_{x}}_{ij} + \nabla \Phi_{ik}^{x} S^{(i)_{x}}_{ik} + \Phi_{ik}^{x} \nabla S^{(i)_{x}}_{ik} - \nabla\Omega_i^x S_i^{x}.
\end{align}

\begin{figure}
\centering
\includegraphics[scale=0.5]{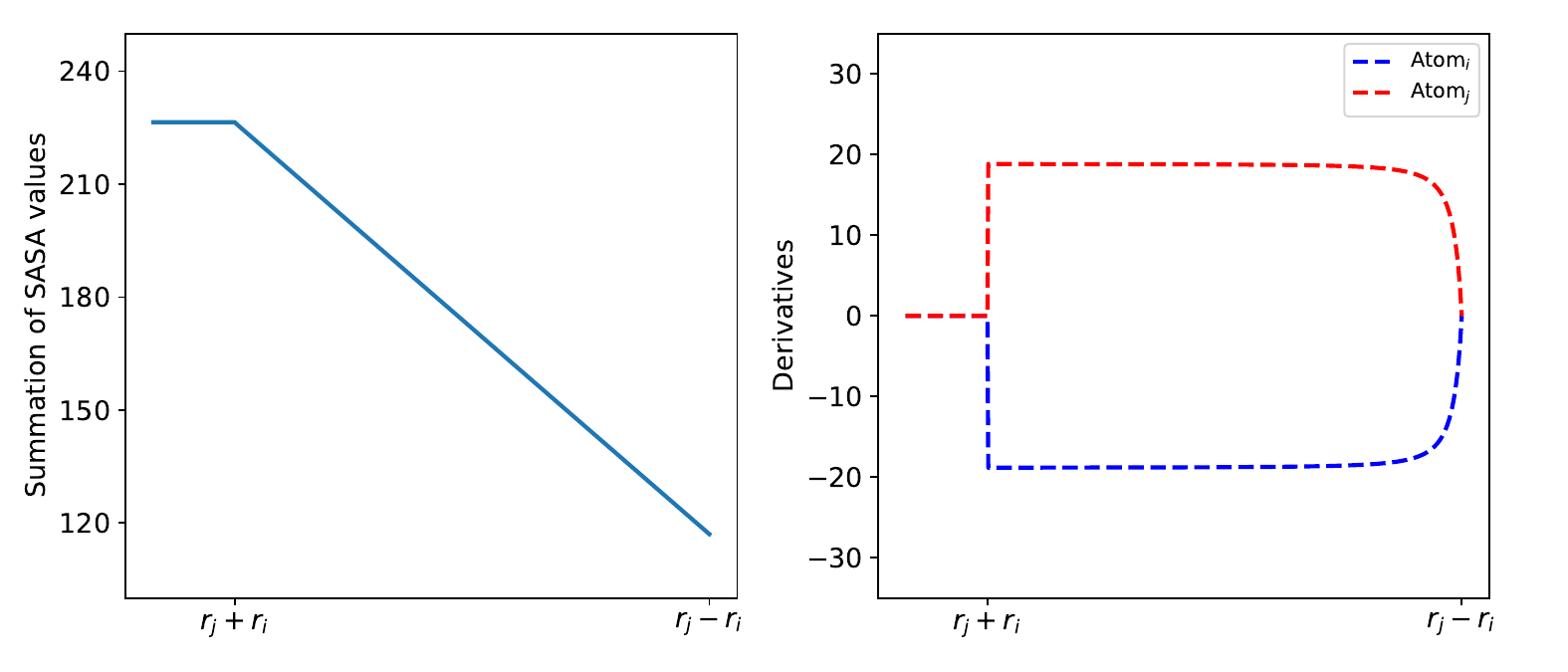}
\caption{Trend of SASA and the magnitude of atomic derivatives as the distance of two atomic centers is changing. The x-axis shows the distance between the atomic centers, decreasing to the right. The summation of SASA values is on the left, while the atomic derivatives are on the right.}
\label{method:sing}
\end{figure}

By substituting Eqs.~\eqref{eqs:vertex}, \eqref{eqs:phiedge}, \eqref{eqs:edge}, \eqref{eqs:trianglev} and \eqref{eqs:triangle} into Eqs.~\eqref{eqs:sainexclu} and \eqref{eqs:inexclu}, we can obtain the surface area and the derivative for every atom. More details of the description and the equations can be found in \cite{hummel18,hummel14} . 
However, the atomic derivatives may become discontinuous when atoms are approaching. Through detailed analysis in \cite{hummel14}, it can be seen that most common type of discontinuities occur when two atoms become externally tangent. We examine here the simplest case where only two atoms with extended radii $r_i=2.9, r_j=3.1$~\AA~are considered to show the possible singularity in the calculation. 
The changing process of the SASA values and the atomic derivatives is given in Figure~\ref{method:sing}. In the left, the summation of SASA values is continuous as two atoms are approaching from a long distance and the minimal value occurs when two atoms are internally tangent. In the right, the magnitude of the atomic derivatives has a leap when two atoms are externally tangent at which the singularity occurs. 
As investigated in \cite{hummel14}, the singularity rarely happens when the solvent radius is 1.4~\AA. In the implementation, we need one threshold value $\epsilon$ to determine the external tangency if $| |p_i - p_j| - (r_i + r_j)| < \epsilon$. When $\epsilon=10^{-6}$, the external tangency was not detected after simulating a few systems for thousands of steps. 
When $\epsilon=10^{-3}$, there may occur one such case after dozens of simulation steps. On the other hand, the magnitude of the leap in the atomic derivatives is less than 20 \AA$^2$ and the surface tension is usually less than 0.01 kcal/(mol$\cdot$ \AA$^2$), so the impact of such singularity is insignificant compared to the overall forces for exterior atoms whose magnitude can be more than 10 kcal/mol.
Given the little chance of happening and the tiny impact of the singularity on the overall forces of the atoms, we can expect that the discontinuity in SASA calculation will have negligible impact on the trajectories in GB/SA simulations.

\subsection{Implementation of dSASA on GPUs}
As the implementation of MD in Amber has been done on GPUs \cite{gotz12, case17}, our main concern here is to implement the method on GPUs to 
speed up the calculation, making it applicable to longer time MD simulations, so we discuss more details about the necessary treatments
on GPUs below. 

\subsubsection{Weighted Delaunay Tetrahedrization}
In the wDT, all possible connections among local points can be detected. Based on the coordinates and weights, four local points are assigned to a tetrahedron having the property that no other points will be inside the circumsphere of this tetrahedron. 
The wDT can be sequentially computed on CPUs with complexity $O(N \log N)$. One parallel algorithm on GPUs to generate exact wDT for a large dataset remains a challenge. The typical GPU implementation of wDT often obtains a near-Delaunay tetrahedrization followed by transformation to CPU to generate the valid wDT. However, an algorithm gReg3D \cite{nanjappa12} which is able to compute the exact wDT on GPUs was implemented a few years ago. 
In the algorithm, all points are first included into a cube with an appropriate size and next the points will be assigned to smaller cubes. The local tetrahedrization for these small cubes are calculated in parallel, followed by checking the consistency and applying necessary modifications to achieve the final wDT. 
The largest size of the initial cube in this algorithm is 512, so the point set with a greater size will be rescaled to fit into the cube, which may introduce a few errors. As the SARS-CoV-2 spike protein containing nearly 4000 residues typically occupies a cube with size 256, this algorithm can compute an exact wDT for proteins with thousands of residues on GPUs.
As shown in Figure~\ref{method:lag1}, the black points are the center of atoms and dashed lines connecting black points denote the edges in the tetrahedrization. The purpose of this step is to obtain the dual complex $\mathcal{C}$ and the conjugated Laguerre diagram. In this diagram, every point
is represented by a cell enclosed by the red solid lines and dashed curves, then the surface area can be calculated with such information.

\subsubsection{Extraction of exterior simplices and calculation of surface areas}
The information of close atoms in the tetrahedra 
has been achieved in the wDT above. However, some unrealistic connections and the interior contacts will not have contribution to surface areas. Then we need to extract the exterior simplices $\partial \mathcal{C}$ containing the feasible vertices, edges and triangles along with the connections among them, which will finally be used to calculate SASA. The extraction of simplices on CPUs is serial and cannot be implemented on GPUs directly. Here we provide more details for the extraction of simplices on GPUs. 

We first need to create lists of unique triangles, edges and vertices with connection information. Take one tetrahedron as an example. It contains 4 triangles as faces, 6 edges and 4 vertices. One element in the list of triangles contains the information of three vertices and the tetrahedra it belongs to, one edge element contains the information of two vertices and the triangles and tetrahedra it is in, while one vertex element has the information of the associated edges and tetrahedra.  

\begin{figure}
\centering
\includegraphics[scale=0.5]{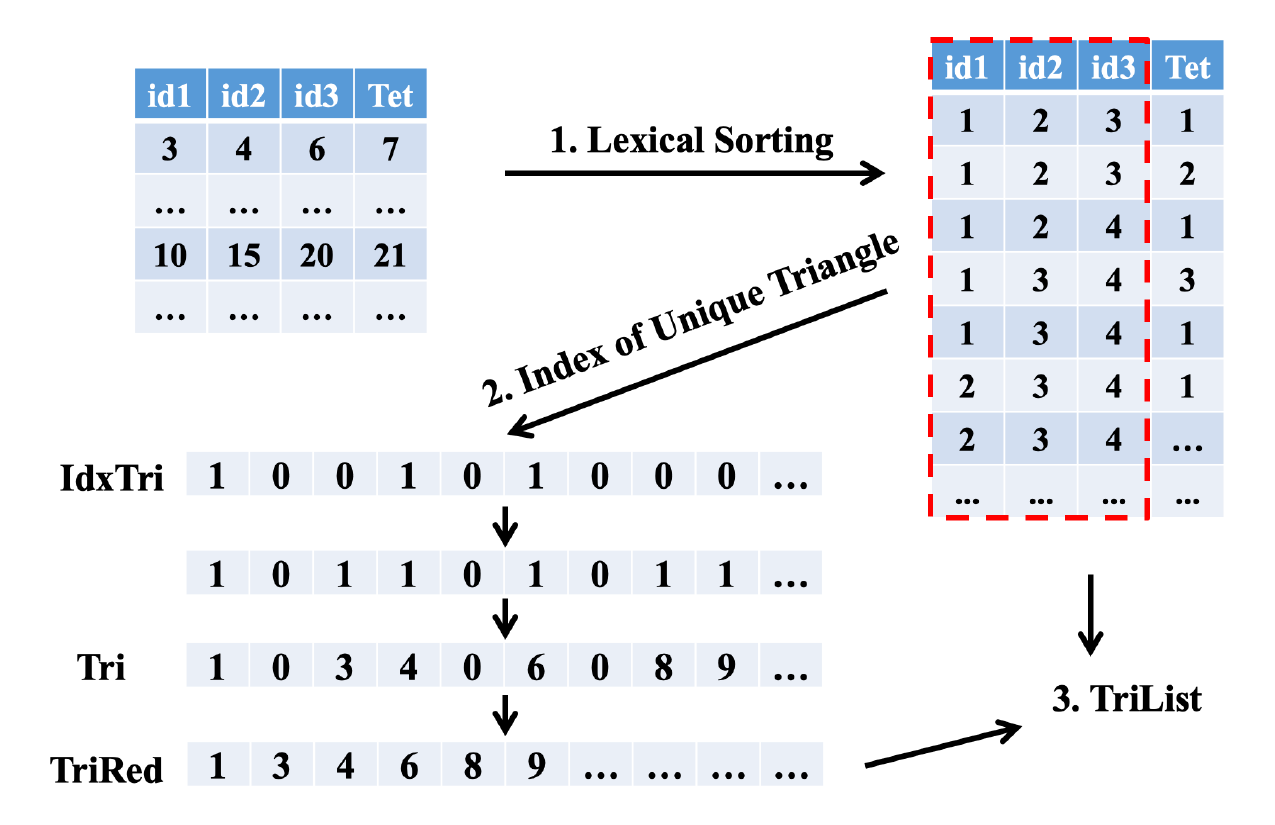}
\caption{Procedures to create the list for unique triangles.}
\label{method:sorting2}
\end{figure}

The exported information from gReg3D provides the vertex indices of all tetrahedra, so we can pull out all the information of edges and triangles in tetrahedra by taking advantage of the parallelism on GPUs. One list of all possible triangles is created and every element has three vertices and the associated tetrahedra. 
The list of all possible edges is constructed with each element containing two vertices and the associated tetrahedra. Suppose there are $N$ tetrahedra, then $4N$ possible triangle elements and $6N$ possible edge elements are in the lists respectively. However, some duplicate elements need to be filtered out to create the lists of unique elements.
Take the list of triangles as an example shown in Figure~\ref{method:sorting2}. We first do lexical sorting for the list of triangles by the first three indices, bringing the same triangles next to each other. Next, with parallel computing, we can identify the number of the associated tetrahedra for a triangle. One triangle can belong to at most two tetrahedra. In the following assignment, every element will check the three vertices of the next element in the list, if three vertices are identical, then we will assign `1' to the newly created array ``idxTri'' which indicates the triangle is duplicate, otherwise we assign `0'. 
We then can identify the unique triangle elements in the list and record the indices in one array ``Tri'': a `1' in ``idxTri'' is a new element, and the second `0' in two consecutive `0's indicates a new element as well. Now we have the indices for all unique triangles and the number of associated tetrahedra, then we can create the data structure ``TriList'' for these triangles. 
When creating triangles, we extract a list of edges and assign the associated triangle indices to this new list, which will provide partial information to edges. Following the similar process above, we can create the lists of unique edges and unique vertices respectively.

With the lists for vertices, edges and triangles, we next can classify them into interior and exterior, and record the indices of exterior simplices. The basic rule is: the triangle with at most one associated tetrahedron is exterior, and one edge is exterior only if its two vertices are both exterior. 
As shown in Figure~\ref{method:lag1}, the atoms in green and blue are exterior while the grey atom $p_3$ is interior whose SASA value will be 0, the edge $p_1 p_2$ is exterior and the edges $p_1 p_3, p_2 p_3$ are interior. 
For a molecule with 1700 atoms, the numbers of tetrahedra, triangles and edges are more than 11200, 22600 and 13000 respectively. These simplices 
can be independently implemented in parallel on a GPU to speed up the classification.
Given the lists of exterior vertices, edges and triangles, we then can calculate atomic surface areas and gradients by Eqs.~\eqref{eqs:sainexclu} and \eqref{eqs:inexclu} in parallel and assign the values to individual atoms.

\subsection{Simulated protein systems, RNA systems}
Trp-cage variant Tc5b (PDB code, 1L2Y \cite{Neidigh02}) has 20 residues with the sequence NLYIQWLKD GGPSSGRPPPS, and the burial of the hydrophobic tryptophan side chain provides a driving force for its folding. It contains an $\alpha$-helix, a $3_{10}$-helix and a C-terminal PP$_{\text{II}}$ helix and the Trp indole ring encapsulated in a cluster of Pro rings. The structure was determined via NMR experiment.

Homeodomain variant (PDB code, 2P6J \cite{Shah07}) contains 52 residues with sequence MKQW
SENVEEKLKEFVKR$\text{H}^\delta$QRITQEEL$\text{H}^\delta$QYAQRLGLNEEAIRQFFEEFEQRK. It is a variant of \textit{Drosophila melanogaster} engrailed homeodomain and was solved by NMR. It consists of three $\alpha$-helices connected by loop regions.

The 14-mer cUUCGg tetraloop hairpin RNA (PDB code, 2KOC \cite{Nozinovic10}) is NMR-solved model and contains 14 bases with sequence GGCACUUCGGUGCC, with the common and highly stable UUCG loop. 

The stem loop C 5`AUA3' triloop of Brome Mosaic virus RNA (PDB code, 1ESH \cite{KimC00}) is composed of 13 bases with sequence GGUGCAUAGCACC. It is designed to contain the triloop AUA in the middle and is an NMR-solved model. 

The CD experiments provide the melting curves for Trp-cage \cite{Neidigh02} and the melting temperature for homeodomain variant \cite{Shah07}. These two protein systems were studied in \cite{Huang18} and the ab initio folding experiments in \cite{nguyen14} using the same force field and solvent model, providing a good reference to quantify the possible improvement by addition of a nonpolar solvation term. As mentioned above, it is important to compare results in which only the component of interest is varied.

\subsection{Details of MD simulation}
In the GB and GB/SA MD simulations, replica exchange molecular dynamics (REMD) was applied to enhance the efficiency of sampling. The setting for the system is as follows: all bonds involving hydrogen were added SHAKE constraints; Langevin dynamics ($ntt=3$) with 1 $ps^{-1}$ collision frequency was used; the time step was 4 fs by following the protocol in \cite{hopkins15} through which the masses of hydrogen atoms can be repartitioned; exchanges between adjacent temperature replicas were attempted every 1 ps; conformations were extracted every 0.1 ns. 
Trp-cage and homeodomain were parametrized by ff14SBonlysc \cite{maier15} with GBNeck2 \cite{nguyen13} and mbondi3 radii \cite{nguyen13}. 
The surface tension parameter was extensively tested in pwSASA \cite{Huang18} for proteins and 5, 7 and 10 cal/(mol$\cdot$ \AA$^2$) provided reasonable results when compared to simulations of a model system in explicit water. Thus we implemented the GB/SA simulations for LCPO, pwSASA and dSASA with 5 or 7 cal/(mol$\cdot$ \AA$^2$).  

For Trp-cage, two production runs starting from unfolded and the first NMR structure were simulated for 1.0 $\mu s$ in GB, LCPO, pwSASA and dSASA, with a REMD ladder of 8 temperatures (247.7, 264.0, 281.4, 300.0, 319.8, 340.9, 363.3, 387.3 K) . The backbone RMSD cutoff 2.0~\AA~was applied to calculate the fraction of folded. For homeodomain variant, two production runs starting from extended and the first NMR structure were simulated with a ladder of 10 temperatures (288.7, 300.0, 311.7, 323.9, 336.6, 349.8, 363.5, 377.7, 392.4, 407.8 K) for 2.0 $\mu s$ in GB and pwSASA, 1.0 $\mu s$ for LCPO and dSASA. The backbone RMSD cutoff 5.0~\AA~was used to calculate the fraction of folded. These cutoffs are consistent with the study in \cite{nguyen14,Huang18}.
MD simulations were not performed for the RNA systems, since weaknesses in current RNA force fields make it challenging to obtain stable simulations for stable hairpins such as 2KOC even with fully explicit water \cite{mrazikova20}. We therefore restrict our analysis to the SASA accuracy for RNA.

\section{Results} \label{sasa:allresults}

\subsection{Molecular SASA estimation}\label{sasa:results}
\begin{figure}
\centering
\includegraphics[scale=0.5]{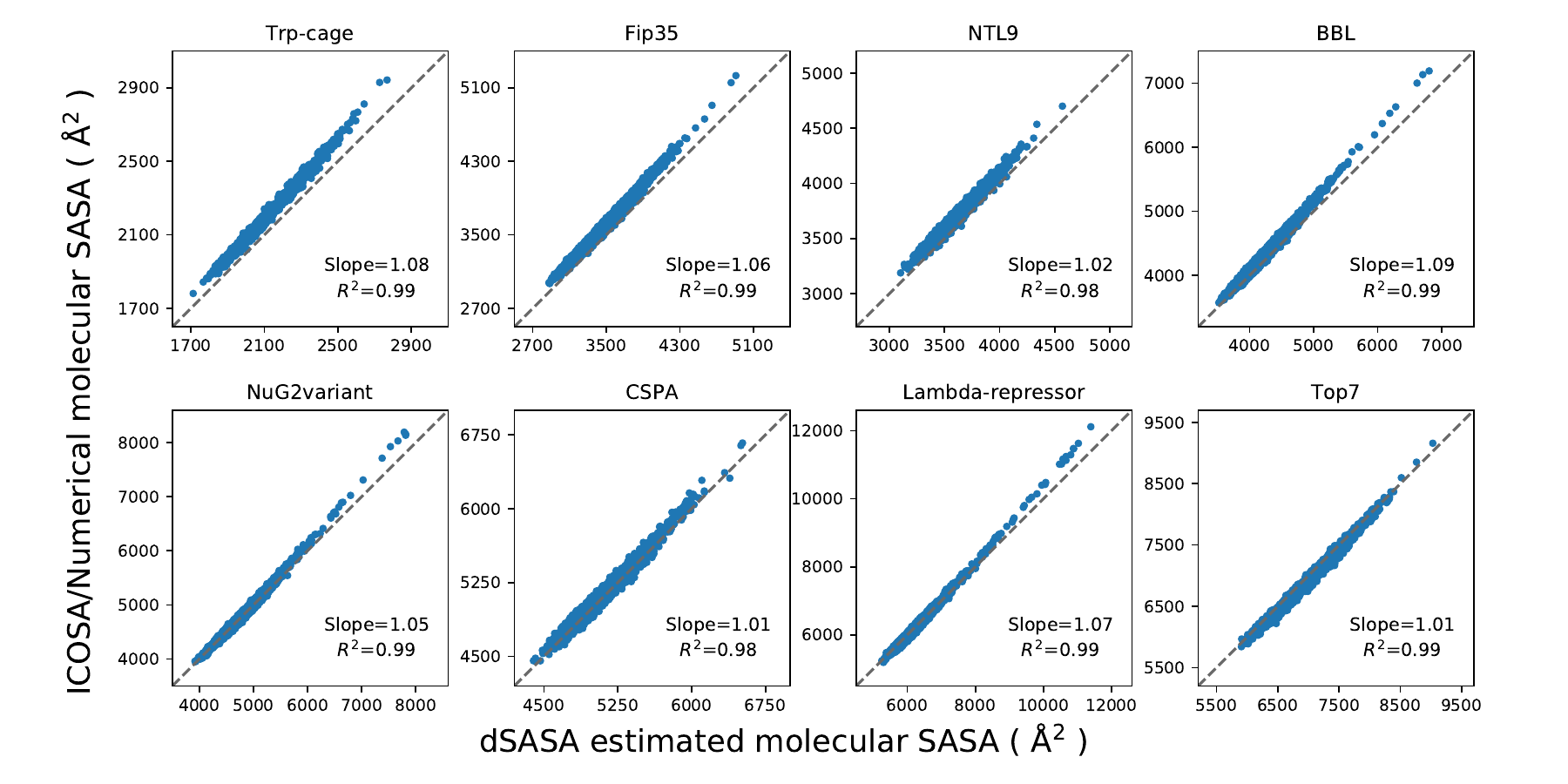}
\caption[Comparison of molecular SASA between dSASA and ICOSA.]{Comparison of molecular SASA between dSASA and ICOSA numerical method for the proteins. Each point represents one conformation of the protein. The diagonal dashed lines indicate perfect agreement.}
\label{results:value1}
\end{figure}

In the test set to validate SASA estimation, we selected eight proteins from the set of the previously examined proteins for ab initio protein folding \cite{nguyen14}, in which the set of proteins have diverse topologies:
Trp-cage (20 residues), Fip35 (33 residues), NTL9 (39 residues), BBL (47 residues), NuG2variant (56 residues), CSPA (69 residues), Lambda-repressor (80 residues) and Top7 (92 residues).
An ensemble of structures for every protein were extracted from the protein folding trajectories in that work with an even interval, obtaining a set of conformations with diverse atomic and molecular SASA values. The corresponding data were estimated for every conformation using ICOSA, LCPO and pwSASA and dSASA.
The numerical method ICOSA can provide more accurate SASA estimation for molecules than LCPO and pwSASA, so we mainly compared the molecular SASA values from dSASA with ICOSA. The results are shown in Figure~\ref{results:value1} and every point represents one conformation in the trajectory. In all the systems, dSASA can generate well-correlated SASA values with ICOSA as Pearson correlation coefficients of the linear regression are $R^2 =0.98, 0.99$ and the slopes are also close to 1. We note that this close agreement in fact demonstrates the relative accuracy of ICOSA as compared with our exact geometric method. However, ICOSA lacks the derivatives needed for MD simulation.

\begin{figure}
\centering
\includegraphics[scale=0.5]{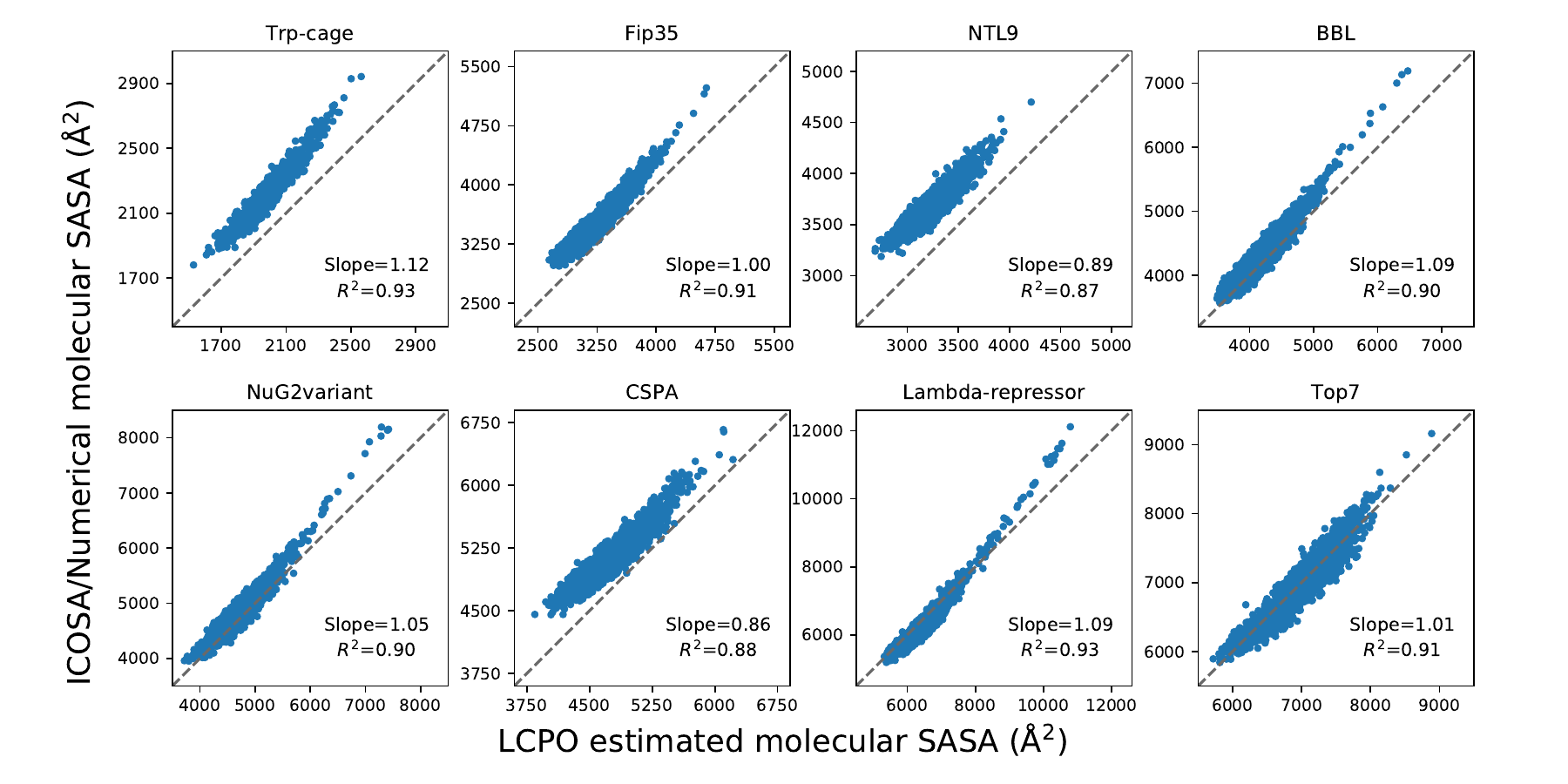}
\caption[Comparison of molecular SASA  between LCPO and ICOSA.]{Comparison of molecular SASA  between LCPO and ICOSA numerical method for the proteins. Each point represents one conformation of the protein. The diagonal dashed lines indicate perfect agreement.}
\label{results:value2}
\end{figure}

\begin{figure}
\centering
\includegraphics[scale=0.5]{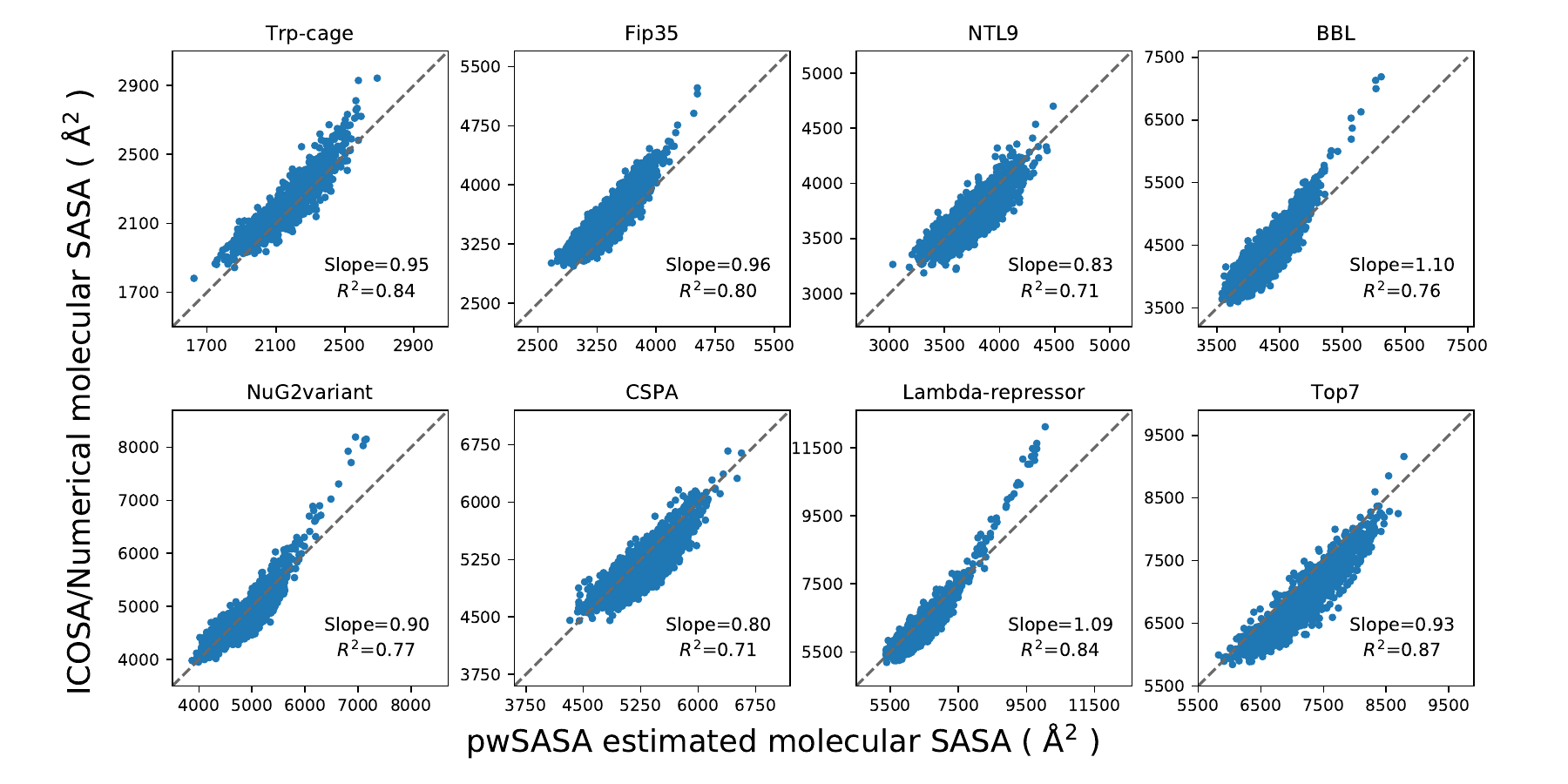}
\caption[Comparison of molecular SASA between pwSASA and ICOSA.]{Comparison of molecular SASA  between pwSASA and ICOSA numerical method for the proteins. Each point represents one conformation of the protein. The diagonal dashed lines indicate perfect agreement.}
\label{results:value3}
\end{figure}

The molecular SASA values from LCPO and pwSASA were calculated using the same conformations and compared with ICOSA method. As shown in Figure~\ref{results:value2}, the discrepancy of LCPO molecular SASA values with ICOSA values varies for different conformations in the systems. Overall, the LCPO tends to underestimate the values with most of the values falling above the perfect agreement line. The correlation coefficients are around 0.9, but the slope and correlation are are lower than 0.9 for NTL9 and CSPA. 
The results became worse with pwSASA shown in Figure~\ref{results:value3}, where  the similar phenomenon arose: the discrepancy of pwSASA molecular SASA values with ICOSA values varies for conformations in the systems; furthermore, all the correlation coefficients are less than 0.9 and the coefficients can be less than 0.8 for NTL9, BBL, NuG2variant and CSPA. This reflects on the goal of pwSASA being speed rather than accuracy.

Both LCPO and pwSASA use pairwise overlaps
which can be relatively faster, but the inexact calculation will lead to inaccuracy especially for some types of atoms. For example, both methods occasionally produce unphysical negative SASA values for individual atoms, while the atomic values from dSASA are all nonnegative.
LCPO includes the higher order correction terms to account for the overlap between two neighbors of the central atom. Then several atom types are defined by the environment, such as the atomic number, the number of bonded neighbors and the state of hybridization. Every atom type has predefined parameters through training. 
Because of less data for certain atom types, the parameter values for these atoms may generate unexpected values which explains the discrepancy for the proteins above. As noticed in \cite{weiser99}, the method works better for more exposed atoms and tends to overestimate the surface area of some buried atoms, such as the carboxyl carbons of the amino acids Asp, Asn, Gln and Glu. pwSASA only considers the overlap between neighboring pairs and compensated by training highly specific parameters applicable only to the atomic environments present in standard amino acids. Even though pwSASA introduced many additional empirical parameters, it suffers similar problems as LCPO, likely due to weak transferability of the pair parameters, and accounting for 3-body and higher terms only in an average way. In fact, the pwSASA values are not the original SASA values because it introduces an empirical adjustment to the SASA values to compensate for a systematic divergence from the numerical ICOSA calculations, which may adversely affect the overall accuracy as well.

\subsection{Speed Comparison in GB/SA MD} \label{sasa:speed}
The original version of dSASA was implemented on CPUs, and the transformation to a GPU version speeds up the computation by taking advantage of the parallel computing. The program written in CUDA was integrated to the Amber20 version. In sander, pmemd, or pmemd.cuda, setting the $gbsa$ flag to 4 in GB suite will activate GB/SA simulations with dSASA. 
We first compared the speed of GB/SA MD on GPU with CPU and next with other methods using 10 proteins with various sizes ranging from 10 residues to 190 residues. 4 of them were selected from above and the PDB code of the rest are given: 10 (PDB code, 5AWL), 20, 33, 52 (2P6J), 69, 92, 110 (1BYW), 130 (1E6K), 162 (2B75) and 190 (1BK7). 

\begin{figure}
\centering
\includegraphics[scale=0.45]{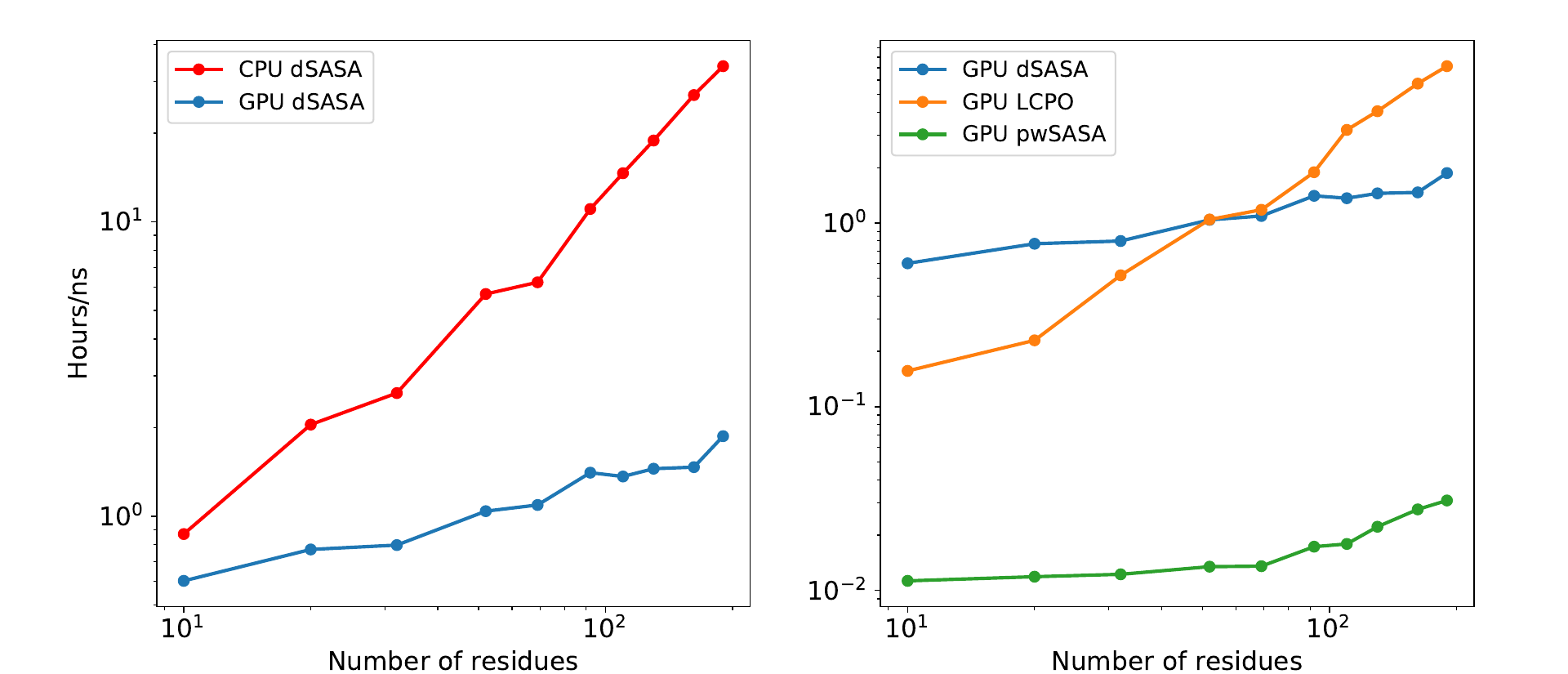}
\caption[Speed comparison in GB/SA.]{Left: Performance of dSASA for various sizes of proteins on CPU and GPU. The y-axis shows the wallclock time in hours to complete 1 ns simulation. Right: Performance of different methods for proteins on GPU. }
\label{results:time1}
\end{figure}

We benchmarked the method on the CPU Intel Platinum 8268 2.90GHz and on the GPU Nvidia RTX 2080 Ti. The estimated wallclock time for one step SASA calculation on CPU is from 12 ms (10 residues) to 90 ms (190 residues), while the time trend on GPU is from 12 ms to 30 ms. The wallclock time here and below is averaged over several simulations. The speed on CPU is comparable to the speed on GPU when the size of protein is small, and the speed on GPU becomes relatively faster as the size of the protein increases. The estimated wallclock time in hours required to complete 1 ns GB/SA simulation in Amber is given in the left of Figure~\ref{results:time1}. 
The wallclock time required to finish 1 ns simulation on the CPU increases from 0.87 hour (10 residues) 
to 33.80 hours (190 residues), while the corresponding trend of timing on the GPU is from 0.6 hour to 
to 1.87 hours. 
When the size of protein is small, the improvement of GPU over CPU version is small with speedup 1.44 times for 10 residues, 
while the speedup grows as the size of molecule is increasing to 
18.07 times for 190 residues. 
For the system with 190 residues, we tested the parallel version with multiple CPUs. The GPU version is still much faster than the parallelized CPU version. The speedup of the GPU version over parallel version with 4, 8, 16 CPUs is 7.6, 9.1, 4.8, respectively.

We next compared the speed with other methods in Amber using the same set of systems above. The wallclock time in hours needed to complete 1 ns simulation is provided in the right of Fig.~\ref{results:time1}. The speed of pwSASA is stable and much faster than other two methods as it is designed to have a fast and simple estimation with more approximations, so we can confirm that the accuracy of the method is at cost of speed as LCPO and dSASA can reproduce more accurate SASA values by the testing above. 
As can be seen in Fig.~\ref{results:time1}, the speed of dSASA is more stable as the size of the system grows, while LCPO has a steeper increasing trend when the size of the molecule becomes larger. When the size of the system is small (10 residues), dSASA is slower than LCPO. While the crossing point is around the size of 69 residues and after that, dSASA starts to gain advantage over LCPO for 69, 92 residues and the speedup becomes greater for larger systems. This phenomenon fits the computation complexity for these two methods: the computation complexity of LCPO is $O(N^2)$ while the complexity of dSASA is $O(N \log N)$ (the complexity of tetrahedrization algorithm), so the system with a larger size will have relatively better performance.
Another reason is that the GB/SA simulation with LCPO is CPU/GPU hybrid so the data transfer further hinders the performance. On the other hand, pwSASA is $\sim$80 times faster than dSASA for small proteins, reducing to $\sim$50 times for larger proteins. While pwSASA is a pairwise method and its computation complexity is also $O(N^2)$, however its computation is efficiently embedded in the overall energy computation which is of the same order; thus pwSASA adds little to the overall computational cost.

\subsection{Stability Analysis of Proteins} \label{sasa:protein}

dSASA can estimate the molecular SASA accurately in proteins with diverse shapes based on the testing above. With the speedup on GPUs, it is now possible to examine its performance in GB/SA simulations and evaluate its effect on the stability of protein structures. However, comparison of the simulation results to experiments should be done with the understanding that deviations can arise from inaccuracy in any of the multiple components of the model (such as the biopolymer force field or GB model, which typically contribute more than the SASA term). Additionally the implicit solvation models may become less reliable away from 300 K. 
Given these caveats, we simulated two proteins with dSASA and compared the results with experiments, GB-only and other SASA methods. The trends in the simulations can provide useful information albeit with these limitations. The thermal stability profiles of Trp-cage and the fraction of folded on homeodomain are calculated from REMD at various temperatures. The temperature ladder for Trp-cage contains 8 replicas (247.7-387.3 K, see Methods for full list). The ladder for homeodomain has 10 replicas (288.7-407.8 K). 

The default surface tension value in Amber GB/SA module is 5 cal/(mol$\cdot$ \AA$^2$) \cite{stikoff94} and the recommended value is 7 cal/(mol$\cdot$ \AA$^2$) \cite{still90} for pwSASA \cite{Huang18}.
For Trp-cage, we tested surface tension values of 5 and 7 to examine the performance of the method. For homeodomain, we only used the value 7, which can produce more near-native states in the trajectories. 

\begin{figure}
\centering
\includegraphics[scale=0.45]{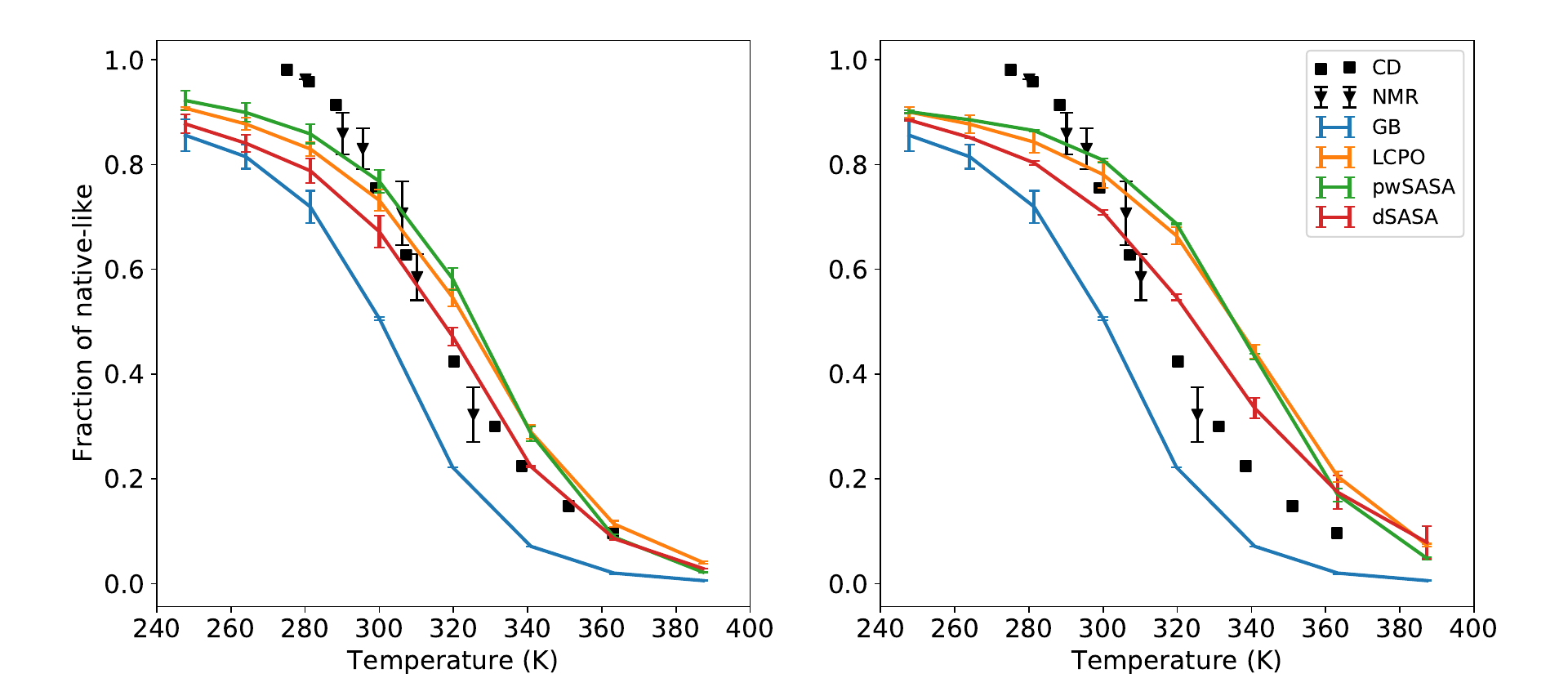}
\caption[Thermal stability profiles for Trp-cage.]{Thermal stability profiles for Trp-cage with surface tension 5 (left) and 7 cal/(mol$\cdot$ \AA$^2$)(right) in GB and GB/SA simulations respectively, including experimental data.}
\label{results:thermal1}
\end{figure}

The thermal stability profiles of Trp-cage with surface tension 5 and 7 cal/(mol$\cdot$ \AA$^2$) were first computed and the fractions of the near native conformations are given in Figure~\ref{results:thermal1} for all methods. 
The left shows the results with the surface tension 5 cal/(mol$\cdot$ \AA$^2$), while the right shows the results with the surface tension 7 cal/(mol$\cdot$ \AA$^2$). With surface tension of 5, all three methods incorporating the nonpolar term achieve better agreement with experiments by creating more stable thermal stability profiles at various temperatures. 
At 300 K, simulations using dSASA generate $\sim$75\% fraction of folded states while other methods with nonpolar term and experimental results are $\sim$80\% and the GB-only leads to $\sim$50\%. 
When the temperature is greater than 310 K, the fraction of folded from dSASA is still slightly smaller than other methods but dSASA can produce closer results to the experiments. The predicted melting temperature $T_{m}$ is 316.8 K and the experimental result is around 317 K, which is better than the GB-only (predicted $T_{m}$ is around 300 K) and other methods (predicted $T_{m}$ is greater than 320 K for LCPO and pwSASA).
Moreover, it is expected that a larger surface tension can generate more native-like conformations through which we can demonstrate that the nonpolar term does not disrupt the system. As shown in the right of Figure~\ref{results:thermal1}, all three methods can generate more native-like structures with surface tension 7 cal/(mol$\cdot$ \AA$^2$).  
To show the extent of dynamics and variation in sampled RMSD, we provide the detailed trajectories at several temperatures for Trp-cage starting from extended and native states with surface tension 5 cal/(mol$\cdot$ \AA$^2$) in Supporting Information.

\begin{figure}
\centering
\includegraphics[scale=0.45]{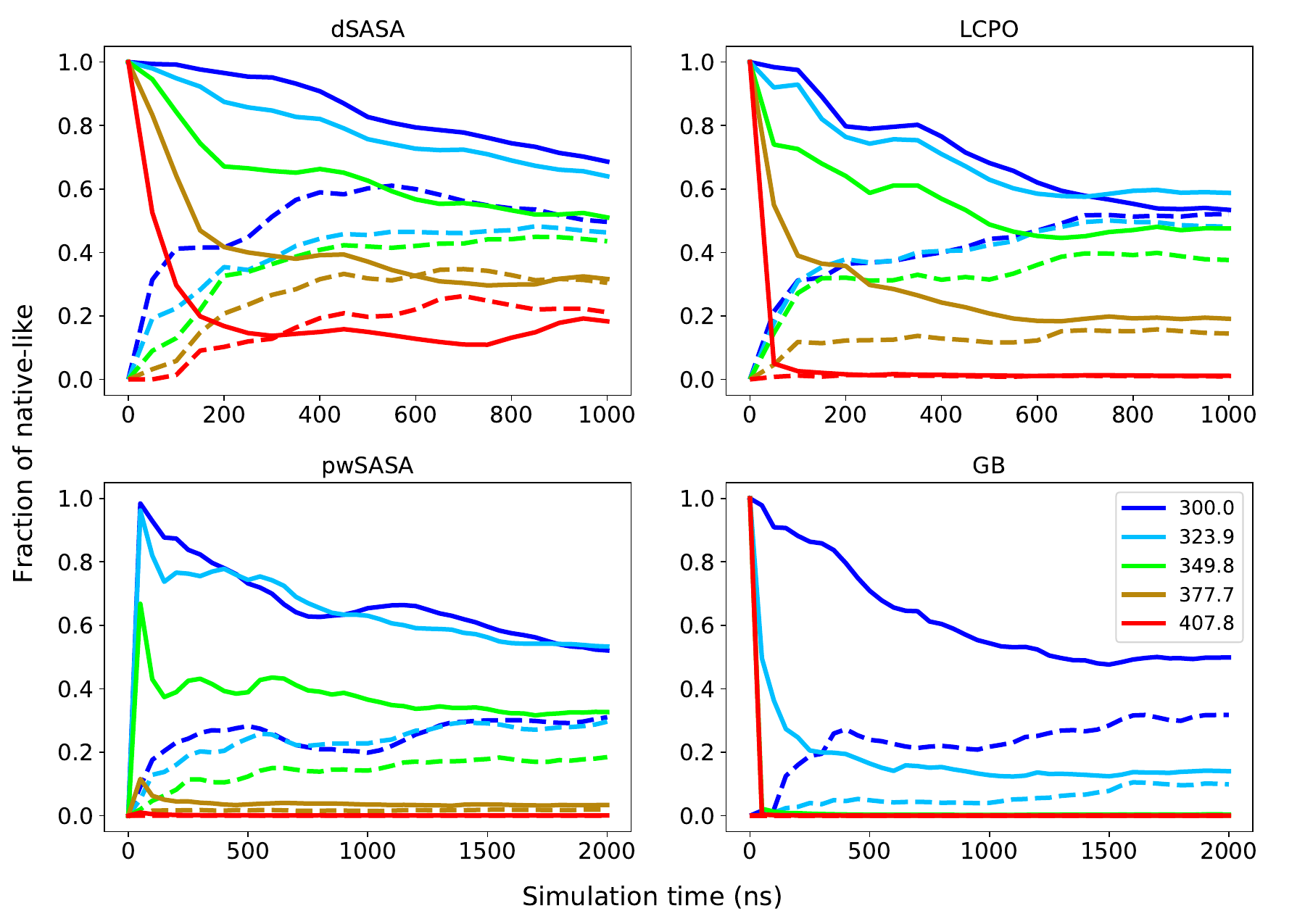}
\caption[Fraction of folded for homeodomain with dSASA, LCPO, pwSASA and GB-only methods.]{Fraction of native-like structures for homeodomain starting from native (solid lines) and extended (dashed lines) at 5 selected temperatures. Top left: dSASA; Top right: LCPO; Bottom left: pwSASA; Bottom right: GB-only.}
\label{results:thermal2}
\end{figure}

For the homeodomain variant, we first carried out REMD with surface tension 5 cal/(mol$\cdot$ \AA$^2$) using pwSASA for 2 $\mu s$, but we noticed that the fraction of folded was far from the experimental results at higher temperatures. The experimental measured $T_m$ is greater than 372 K, so we expect a larger surface tension can achieve better agreement with experiments. We then simulated with surface tension 7 cal/(mol$\cdot$ \AA$^2$) for all GB/SA methods.
All three methods with nonpolar term can improve the fraction of folded at various temperatures compared to GB-only. The fractions of native-like starting from native and unfolded states at selected temperatures are shown in Figure~\ref{results:thermal2}. 
Since the experimental $T_m$ is greater than 372 K, we expect that the system will not collapse at high temperatures. 
The two more accurate SASA methods, LCPO and dSASA, produce more native-like conformations at higher temperatures as shown in Figure~\ref{results:thermal2}. Moreover, dSASA generated more native-like conformations than LCPO at various temperatures, demonstrating that our method can have a better effect on stabilizing the system. 
The detailed trajectories of homeodomain at selected temperatures starting from extended and native states with surface tension 7 cal/(mol$\cdot$ \AA$^2$) are given in Supporting Information. 
Furthermore, at the temperature 300 K, the difference between the native-like fractions of the simulations starting from native and extended is around 20\% for dSASA and LCPO after 500 ns simulations, while that difference is around 40\% for GB and pwSASA even after 1000 ns simulations, indicating that the inclusion of an accurate SASA term might lead to improved sampling and faster convergence (or conversely, that the SASA approximations may lead to increased kinetic traps). 
The improved performance from dSASA still has variance from the experimental results, for example the fraction of native-like at 377.7 K is around 35\% while it is 50\% in experiment. Such discrepancy may come from limitations of the SASA-based nonpolar term, or from the force field and the polar GB term; comparison of simulation results to experiment typically does not reveal which component is responsible, and further model system development is needed.

\subsection{Performance on RNA systems} \label{sasa:rna}
Another advantage of dSASA over pwSASA is that it can be applied to other types of systems such as nucleic acids or small-molecule ligands,
because only the radii of the atoms are considered without defining diverse atom types. Here we test this aspect using two RNA systems, 2KOC and 1ESH, and compare the molecular SASA values with ICOSA and other methods to see if the excellent performance on proteins is transferable to other biopolymers. Simulation results for RNA are not presented in details, since accurate force fields and implicit solvent models for RNA are still a matter of active research \cite{mrazikova20, liebl23}.

\begin{figure}[ht]
\centering
\includegraphics[scale=0.52]{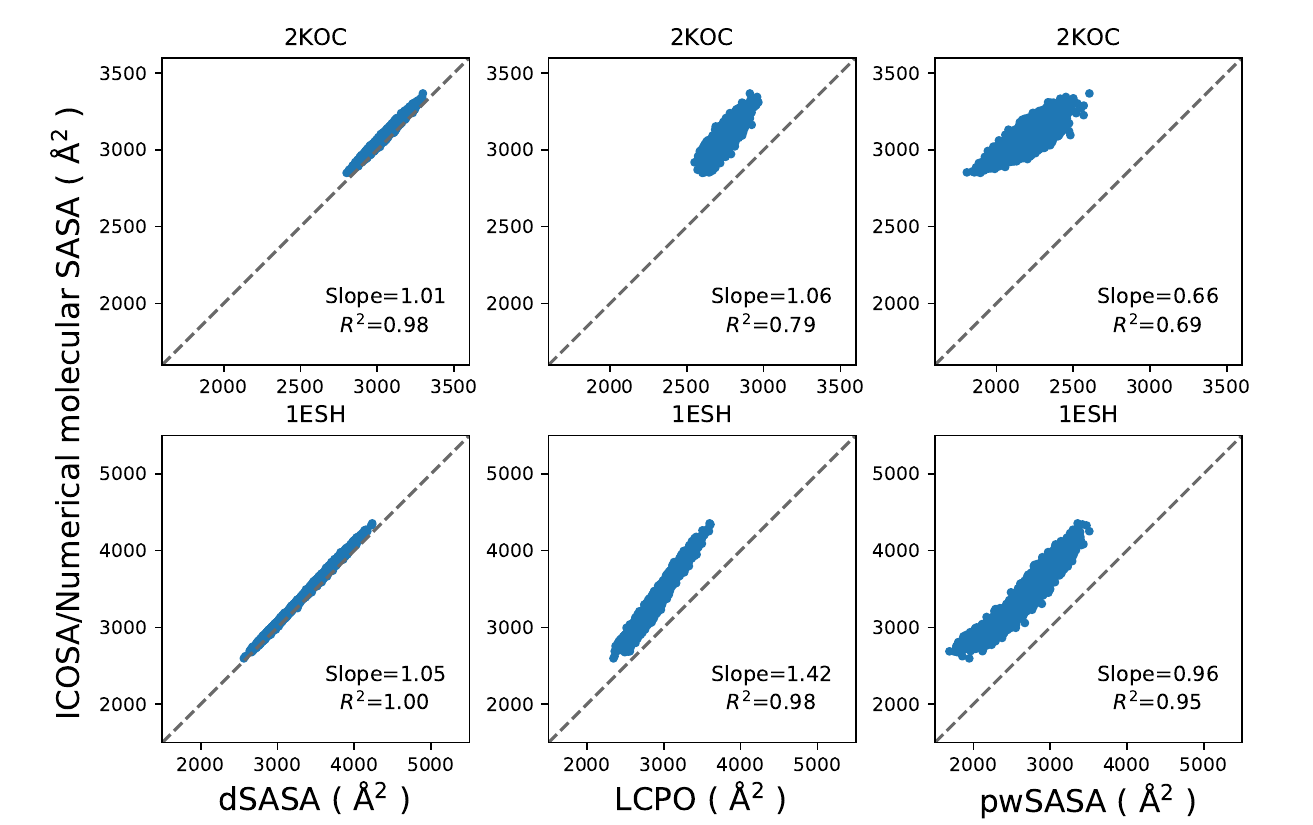}
\caption{Comparison of molecular SASA values from various methods; each point represents one conformation of the RNA. The first row is for 2KOC and the second is for 1ESH. The diagonal dashed lines indicate perfect agreement.}
\label{results:rnaf}
\end{figure}

To validate the SASA estimation, we generated alternate conformations for these RNA systems with the GB-only method for 1 $\mu s$ and extracted 10$^4$ conformations for each system respectively. The comparison of the molecular SASA values between all methods is shown in Figure~\ref{results:rnaf}. dSASA can reproduce SASA values from ICOSA with more than 98\% accuracy for both systems while LCPO and pwSASA produced divergent results for two systems and they had overall underestimation of the SASA values. 
The SASA estimates from pwSASA have large discrepancy from ICOSA method, likely 
because it was trained exclusively on protein data so that the parameters do not accurately model the specific atom types and local environments present in RNA. Overall, the results confirm the expected outcome that dSASA accuracy is transferable across diverse molecular systems.

\section{Conclusions}\label{sasa:con}
In this work, we present dSASA, an analytical SASA evaluation method for molecules and its implementation on GPUs.
In this approach, the weighted Delaunay tetrahedrization is first computed with the atomic coordinates and radii. Next the atomic and molecular SASA values are estimated using inclusion-exclusion method based on exterior simplices from tetrahedrization, resulting in more accurate estimation than LCPO and pwSASA in reproducing numerical ICOSA SASA values. 
The pre-trained parameters for atoms in LCPO and pwSASA can speed up the calculation, but these parameters can have low transferability. This is confirmed by comparison of computed molecular SASA values, which for these approximate methods had smaller correlation coefficients with the values from ICOSA method, such as in the protein systems CSPA and NuG2variant and RNAs. In contrast, estimation from dSASA can reproduce ICOSA values with correlation coefficients greater than 98\%.
Moreover, the current GPU version speeds up the calculation substantially compared with the CPU version, making it applicable for longer time GB/SA MD simulations especially for larger systems. The gain of speed arises from the complexity of the algorithm itself which will scales well for larger systems and from the full implementation on GPU devices removing the data transfer with CPUs.
dSASA has a stable trend of speed as the size of molecule increases but it is still relatively slower than the highly approximate pwSASA. However, dSASA started to outperform LCPO when the system contains around 70 residues, and the performance for larger molecules would be more stable than LCPO because the computation in LCPO is pairwise and its implementation is CPU/GPU hybrid in Amber. 

In the GB/SA simulations, two proteins (Trp-cage and homeodomain) were simulated and compared to other methods and experimental results. dSASA achieved comparable performance with LCPO and pwSASA on the small protein, and the performance became better for the larger system. 
The simulated melting curve with nonpolar term for Trp-cage was more consistent with the experimental measures compared with that without nonpolar term.  
In homeodomain variant, the melting temperature is greater than 372 K so we expect the system can maintain some amount of native-like conformations at higher temperatures. 
The simulations generated more extended conformations with other methods at high temperatures while dSASA can produce more native-like conformations in the trajectories which is closer to the experimental data. However, these comparisons to experiment must be considered in the context of the other force field components as well.

As the program has been rigorously examined, we anticipate to extend its applications to other functions in Amber software, such as post-processing of the trajectories in MM-PBSA or MM-GBSA. The calculation in dSASA is geometry based, obtaining accurate results with the provided atomic coordinates and radii. Therefore it can be used for diverse types of systems without restrictions, such as the simulations of larger proteins, protein-ligand complexes, and protein-nucleic acids complexes. 1 nanosecond GB/SA simulation for 200-residue proteins with dSASA takes around two hours. Even though it is slower than pwSASA, the accuracy of the method will allow us to explore the impact of the nonpolar term on the simulations in the future.
Given the accuracy of dSASA, the calculation of atomic and molecular SASA values can be benchmark data set for the training of the parameters in pwSASA approach for RNAs in the future, which will combine the advantage of its speed on GPUs and the accurate calculation from dSASA. The SASA is certainly an approximation to nonpolar solvation, but many studies have shown that including it improves agreement with experiment for things like protein stability or binding affinities\cite{genheden15}. In addition, dSASA, a more accurate SASA with derivatives implemented on GPUs, will also enable fast polar solvation methods for MD.
Furthermore, the calculation of SASA is based on the diagram of Laguerre intersection cells, so it can be easily extended to the computation of molecular volumes along with the corresponding atomic derivatives. Inclusion of the volume term will provide a more complete description for the nonpolar solvation term. 
With the volume derivatives, we can further examine the impact of the term on the stability of molecules in the MD simulations. Moreover, the current algorithm being used for weighted Delaunay Tetrahedrization, gReg3D, is designed to work on a set of random points and the size of the workspace depends on the distribution of points in the workspace. As it consumes nearly 70\% of wallclock time of our surface area calculation, improvement on this algorithm will further speed up the simulations. 
The program of dSASA written in CUDA will be freely available from the authors.

\section{Associated Content}
Supporting Information available.

Examination of energy conservation after including dSASA term in MD. Detailed trajectories with dSASA, GB, LCPO and pwSASA for Trp-cage and Homeodomain at selected temperatures. Fractions of folded for the second test set of Homeodomain.

\begin{acknowledgement}
This work was partly supported by NIH Grants RM1 GM135136 and R01 GM107104, and by NSF Grant DMS 2054251. We gratefully acknowledge support from Laufer Center for Physical and Quantitative Biology.
\end{acknowledgement}

\bibliography{sasarefs}

\end{document}